\documentclass[a4paper,12pt]{article}
\usepackage{graphicx}
\usepackage{setspace}
\usepackage{vmargin}
\setpapersize{A4}
\setmarginsrb{20mm}{15mm}{20mm}{15mm}{14.5pt}{8mm}{0pt}{11mm}

\begin{document}

\title{A New Artificial Dielectric Metamaterial and its Application as a THz Anti-Reflection Coating}

\author{Zhang~J.$^1$\thanks{E-mail:
jin.zhang@astro.cf.ac.uk}, Ade~P.~A.~R.$^1$, Mauskopf~P.$^1$,
Moncelsi~L.$^1$, Savini~G.$^2$,
Whitehouse~N.$^1$\vspace{0.5cm}\\$^{1}$ School of Physics and
Astronomy, Cardiff University, The Parade,\\CF24 3AA Cardiff,
Wales, UK\\$^{2}$ Optical Science Laboratory, Physics and
Astronomy Department,\\University College London, Gower Street,
London WC1E 6BT, UK}

\date{}
\maketitle

\begin{abstract}

We describe a novel artificial dielectric material which has
applications at millimetre and submillimetre wavelengths. The
material is manufactured from layers of metal mesh patterned onto
thin polypropylene sheets which are then bonded together using a
hot pressing process to provide planar rugged discs which can be
reliably cycled to cryogenic temperatures. The refractive index of
this material can be tuned by adjusting the geometry and spacing
of the metal-mesh layers. We demonstrate its usage by designing
and characterising a broadband anti-reflection coating for a Z-cut
crystalline Quartz plate. The coating was fabricated and applied
to the quartz using the hot press technique and characterized
using a Fourier Transform Spectrometer. The performance is shown
to be in good agreement with HFSS and transmission line modelling
results.

\end{abstract}

\section{Introduction}

Metal mesh technology has been adopted in the millimetre and
sub-millimetre wavelength range (THz frequencies) as the standard
method for fabricating optical filters, beam-splitters and
dichroics \cite{1,2,3,4}. Components fabricated using this
technique have been used in a number of THz cameras and
spectrometers \cite{5,6,7,8,9}. More recently, the same technique
has been used to manufacture half-wave plates using aligned
metallic patterns with air-gap spacing between the layers
\cite{10}. These traditional metal mesh components are not
considered metamaterials because their electromagnetic properties
are not independent of their thickness, i.e. they cannot be
characterized as having a bulk electric permittivity and magnetic
permeability.

This paper demonstrates that closely spaced multiple layers of
metal-mesh films embedded in polypropylene behave as an artificial
dielectric metamaterial. By changing the metal mesh geometry the
effective impedance of the mesh layer is modified changing the
apparent refractive index of the overall material. In Section 2 we
present a theoretical analysis to obtain a physical insight into
this behaviour and to show how the geometrical parameters modify
the effective dielectric constant. This modelling uses a
commercial finite-element analysis package: High Frequency
Structure Simulator (HFSS; \cite{11}), which provides an exact
solution to Maxwell$'$s equations for a unit cell of the
structure.

As a demonstration, a prototype broadband anti-reflection coating
was designed and applied to a single Z-cut quartz plate. In
Section 3 spectral measurements of both the anti-reflection
coating and of the quartz plate before and after coating with the
artificial dielectric are presented.  These results are compared
with predictions from HFSS and our own transmission line model.

\section{Theory and modeling}

\subsection{Theoretical elements}

The metal mesh technology described in Ade et al. \cite{1} is
based on standard two port microwave circuit analysis. The metal
mesh layer is modelled as an equivalent lumped LC filter and the
spacing between meshes is modelled as a transmission line with
impedance, $Z = Z_0/n$ where $Z_0$ is the impedance of free space
and $n$ is the index of refraction of the material between meshes.
The lumped element impedance for a single sheet containing
periodic metal structures has been previously determined by a
number of authors \cite{12,13,14,15} to be dependent only on the
geometric properties of the metal.

For filters and other components built up of many individual metal
meshes, the metallized layers are stacked together with plane
parallel spacers, which consist of either air gaps or dielectric
discs. The model is successful in designing and predicting the
performance when the spacing between layers $d$, is large enough
so that there is relatively little inductive or capacitive
coupling between layers. Air gap devices require an annular
support ring, while the dielectric spacers and metallized sheets
can be fused together using a hot pressing technique to make a
solid self-supporting disc. While the design of these components
is based on radiation at normal incidence to the metal mesh
layers, their performance, which is measured in a converging beam
of half angle 9.6 degrees (see later), agree well. This indicates
that there is not a strong dependence with off-axis angle.

The simplest metal mesh pattern used for low pass filters is
termed a capacitive grid as described in Ade et al. \cite{1}. Here
we describe the properties of a stack of identical capacitive mesh
grids with spacing between the grids, $d << \lambda$ immersed in a
dielectric substrate layer using hot-pressed technology. In these
structures, the two port microwave circuit model breaks down due
to the capacitive coupling between layers. However, the properties
of the components are well described by an effective bulk
permittivity, $\epsilon_r(\nu) > 1$ and bulk permeability, $\mu_r
= 1$ corresponding to a material with effective index of
refraction, $n_{\rm eff} > 1$. By changing the basic mesh
geometry, the mesh layer spacing and the number of mesh layers the
effective permittivity of the newly created artificial dielectric
slab can be modified. This is expected since the lumped
capacitance of the grids will increase the capacitance per unit
length for an electric field parallel to grids, thus increasing
the effective permittivity of the material. A full EM analysis of
structures made of solid metal bars with gaps between them has
been reported \cite{16}. We perform a semi-empirical analysis of
our structures based on 3-D EM simulations using HFSS and compare
these results to measurements of fabricated components.

\subsection{Essential parameters in the build}

Figure 1 shows the basic structure of the capacitive metal mesh
layers embedded in a base dielectric material such as
polypropylene.

\begin{figure}[h!]\label{Fig-1}
\centering
\includegraphics[width=12cm,height=7.5cm,angle=0]{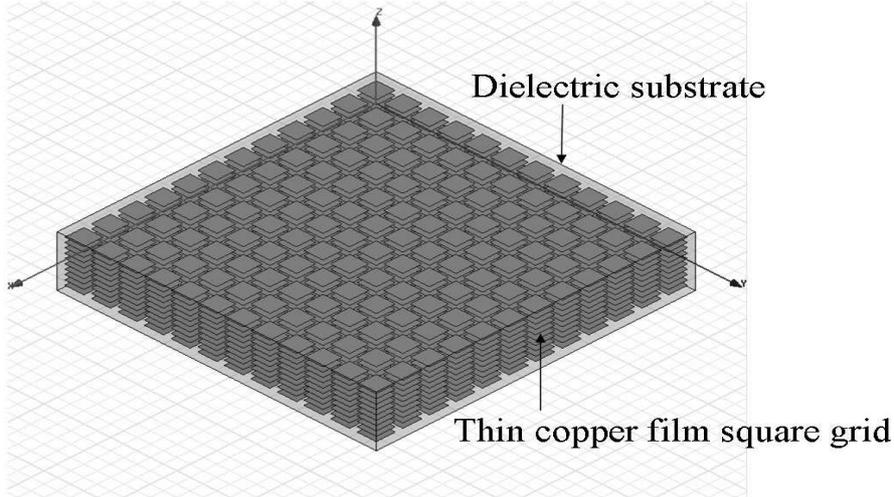}
\caption{The basic structure of the artificial dielectric. The
structure shown has all of the metal meshes in the stack aligned
with each other. Small offsets in alignment from one layer to
another have a small effect on the effective index of refraction.}
\end{figure}

The structure of a single mesh layer is periodic so there are only
three essential geometrical parameters in the model; the spacing
between the layers, \textit{d}, the gap between two adjacent
metallic squares, 2\textit{a}, and the period of the repetitive
square structure, \textit{g}, as shown in Figure 2. It was also
found useful to categorize the mesh patterns by the ratio
\textit{a/g} since for a given grid period this characterizes the
basic frequency dependence of the single mesh.

\begin{figure}[h!]\label{Fig-2}
\centering
\includegraphics[width=12cm,height=8cm,angle=0]{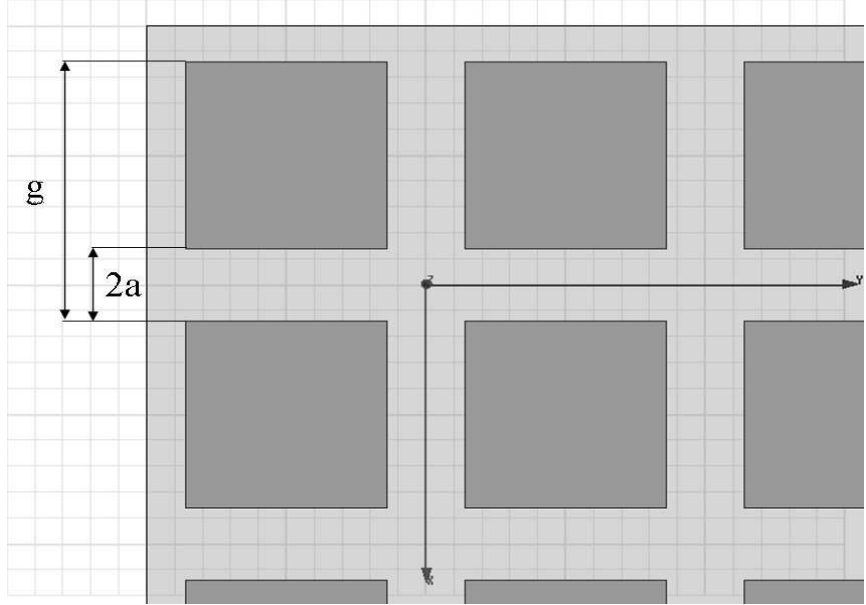}
\caption{Top view of single layer metal mesh pattern.  The
evaporated Cu squares created photolithographically are typically
0.1\,$\mu$m thick.}
\end{figure}

\subsection{HFSS simulation result for metal mesh dielectric}

We used HFSS simulations to explore the optical properties of
these capacitive grid stacks with different geometrical
properties. For these simulations we fixed the basic grid
periodicity, \textit{g}, to be 100\,$\mu$m.  Changing this
dimension along with the spacing between the layers scales the
observed properties to different spectral regions. The simulations
were made for a range of \textit{a/g} ratios from 0.05 up to 0.28
(0.05, 0.1, 0.14, 0.28). For each \textit{a/g}, we also varied the
number of layers (5, 10, 20 layers) and explored the dependence on
the grid spacing, \textit{d}, between adjacent layers (4, 5, 8,
10, 12, 15 and 20\,$\mu$m).

The simulation transmission results for one set of a ten layer
device with \textit{a/g} = 0.28 for various grid spacing are shown
in Figure 3.

\begin{figure}[h!]\label{Fig-3}
\centering
\includegraphics[width=12cm,height=8cm,angle=0]{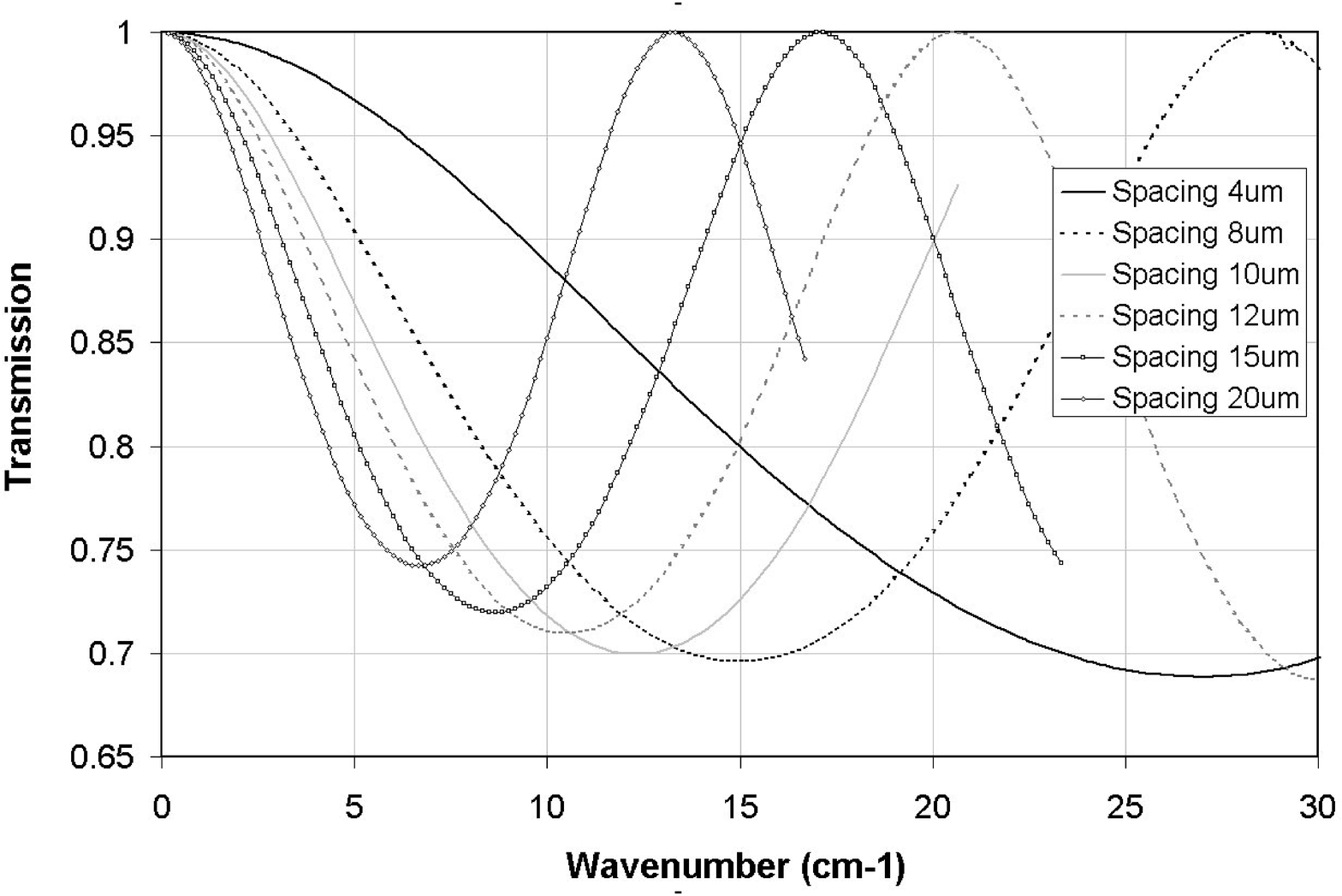}
\caption{HFSS simulated transmission for 10 mesh layers immersed
in polypropylene with fixed grid geometry \textit{a/g} = 0.28 and
\textit{g} = 100\,$\mu$m for mesh layer spacings of 4 (rightmost
minimum), 8, 10, 12, 15 and 20\,$\mu$m (leftmost minimum).}
\end{figure}

The refractive index, \textit{n}, is derived from the transmission
data by assuming that the resultant material behaves like a plane
parallel dielectric plate with thickness, $\Delta x$. From the
HFSS simulation we obtain the complex reflection and transmission
coefficients, \textit{r} and \textit{t}, of the material and
invert the standard Fabry-Perot (FP) transmission formulas
\cite{17}:

\begin{equation}
t = \left[{\rm cos}(2 \pi n \tilde{\sigma}\Delta x)-\frac{i}{2}\left(z+\frac{1}{z}\right){\rm sin}(2 \pi n \tilde{\sigma}\Delta x)\right]^{-1}
\end{equation}

and

\begin{equation}
r = \frac{i t}{2}\left(z - \frac{1}{z}\right){\rm sin}(2 \pi n \tilde{\sigma}\Delta x)
\end{equation}

to determine the index of refraction, \textit{n} and material
impedance, \textit{z}. From these we can compute values for the
relative permittivity, $\epsilon(\tilde{\sigma})$ and
permeability, $\mu(\tilde{\sigma})$. We find for all structures
that $\mu(\tilde{\sigma}) \approx 1$ independent of frequency so
the material can be characterized as a simple dielectric and the
index can be calculated from the FP intensity formula \cite{18}:

\begin{equation}
T (\tilde{\sigma}) = t^2 = \frac{1}{1+[2r/(1-r^2)]^2\,{\rm sin}
^2(2 \pi n \tilde{\sigma} \Delta x)}
\end{equation}

or simply by measuring the transmitted intensity at the first
minimum:

\begin{equation}
T_{min} = \frac{4n^2}{(n^2+1)^2}.
\end{equation}

Inversion of this for the transmission at the first minima
observed in Figure 3 gives the effective refractive index for the
artificial material at that frequency.

\begin{figure}[h!]\label{Fig-4}
\centering
\includegraphics[width=12cm,height=8cm,angle=0]{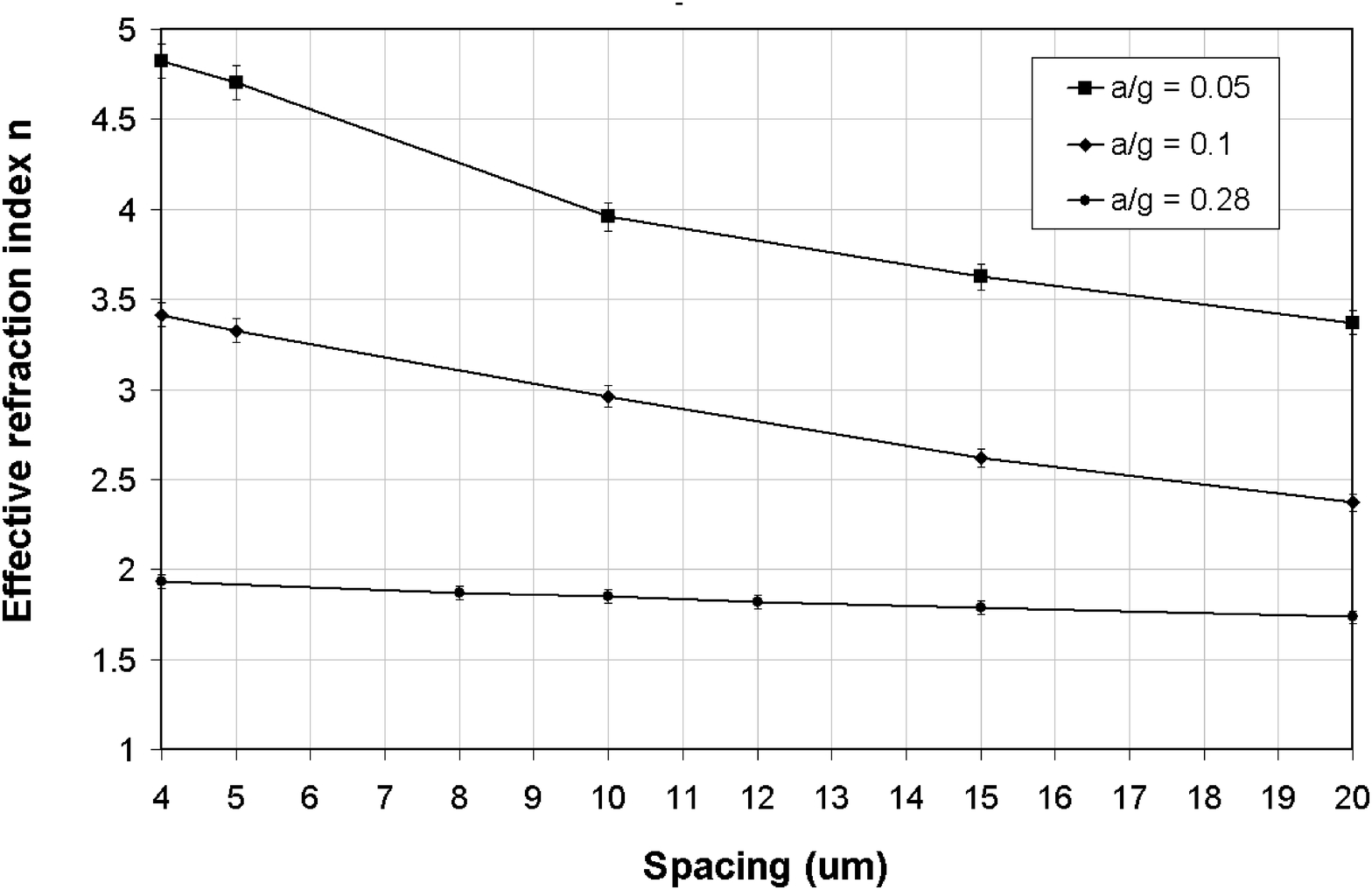}
\caption{Predicted refractive index \textit{n} for a 10 mesh layer
device as a function of mesh layer spacing, for mesh geometries
\textit{a/g} of 0.05, 0.1 and 0.28, at 5\,cm$^{-1}$ wavenumbers
(frequency 150\,GHz), with \textit{g} = 100\,$\mu$m. Errors are
2\,\% due to simulation accuracy.}
\end{figure}

The estimated refractive index as a function of the grid spacing
for each \textit{a/g} ratio is presented in Figure 4. This shows
that for a fixed \textit{a/g} ratio, the refractive index of the
material decreases as the grid spacing increases. It can also be
seen that for a fixed grid spacing, a higher \textit{a/g} ratio
yields a lower refractive index. For both cases an increase in the
grid spacing or the \textit{a/g} ratio results in a decrease per
unit length of the effective artificial material capacitance for
an E-field parallel to grids and hence a decrease in refractive
index. We perform simulations of structures with different numbers
of mesh layers and find that the effective index of refraction is
independent of component thickness and therefore can be considered
a bulk material parameter. Finally, Figure 5 shows that the index
of refraction is weakly dependent on the frequency and that the
metamaterial behaves as an artificial dielectric over a wide range
of wavelengths, corresponding to $g < \lambda$.

\begin{figure}[h!]\label{Fig-5}
\centering
\includegraphics[width=12cm,height=8cm,angle=0]{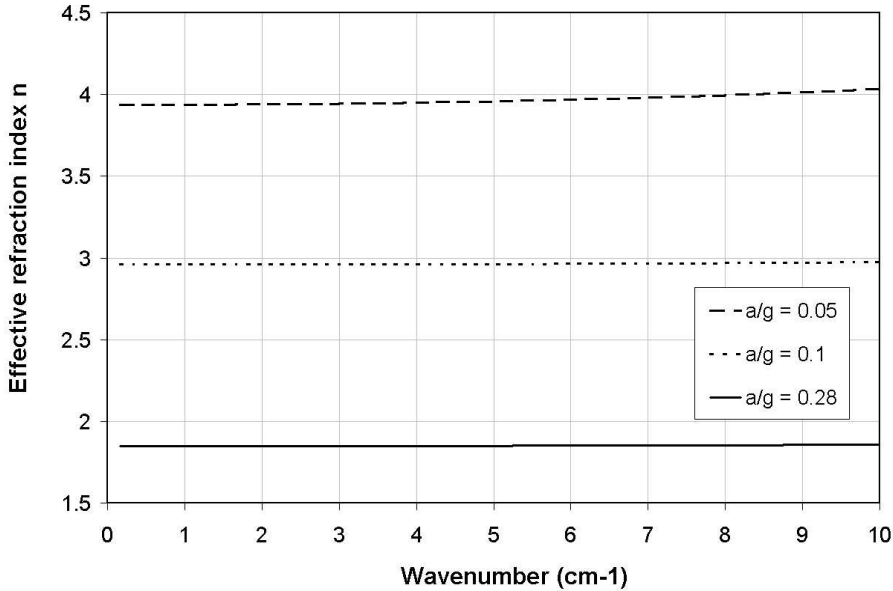}
\caption{Predicted refractive index \textit{n} for 10 mesh layer
device and fixed spacing 10\,$\mu$m, as a function of frequency,
for mesh geometries \textit{a/g} = 0.05, \textit{a/g} =
0.1,\textit{a/g} = 0.28, with \textit{g} = 100\,$\mu$m. In
general, there is a slight increase in the effective index of
refraction with frequency due to the increase in spacing between
the layers relative to the wavelength.}
\end{figure}

\subsection{HFSS model parameters for anti-reflection coating}

One important application of this artificial dielectric is as an
element in a broadband multilayer anti-reflection coating (ARCs).
In this application there are two fundamental material
requirements that need to be met. First, the materials used in
ARCs must have a range of appropriate refractive indices and high
transparency over the desired spectral band.  Secondly, the
material must be mechanically suitable for applying to crystalline
materials.

ARCs are usually used to maximize the device transmission over
spectral bandwidths approaching 100\,\%. Previous designs used
specially prepared polypropylene layers\cite{19,20} that are
loaded with high refractive index powders (TiO$_2$) or ceramic
based materials (e.g., Rogers TMM material \cite{21}) to create a
particular refractive index. These ARCs needed three layers to
achieve a flat response across the band. Each layer required hot
pressing or grinding to the required thickness before being bonded
to the front and back surfaces of the crystalline plates
\cite{19,22}.  Major problems were encountered with both the
previous approaches. The loaded powder layers were slow to
manufacture since the powder needed to be uniformly mixed in the
polypropylene and then the layers had to be hot pressed to the
appropriate thickness.  The ceramics are brittle and could only be
thinned by using a grinding technique which proved to be time
consuming and difficult to handle in thin sheet form.

Compared with the previous techniques, the artificial dielectric
design has the advantages of simple refractive index tuning
through the geometry and spacing of the grids. Without high index
powders added into the material, the metal mesh design has a
complete control over the thickness of the coating layer and the
coating material is not brittle and has better performance in
thermal cycling. Furthermore, since the base dielectric is
polypropylene, it also has low absorption for frequencies up to 20
THz \cite{23}. Thus the artificial dielectric meets with the ARC
fundamental requirements.

To demonstrate the new material we have prototyped an ARC for a
Z-cut crystal quartz plate with refractive index \textit{n}=2.1.
To increase the spectral bandwidth of the low reflection region we
needed two materials with intermediate refractive index values
close to 1.3 and 1.7.  The first of these is realised using
readily available porous PTFE whilst the latter requires the
customised artificial dielectric.

The artificial dielectric is constituted by two essential
elements, a metal mesh grid and the embedding dielectric. The
first is obtained by photo-lithographic techniques adopted in the
manufacturing of far-infrared low-pass filters \cite{24}. The
copper deposited patterns on thin substrates (8\,$\mu$m) of
polypropylene are then inserted into a stack (see Figure 6). The
two metal mesh layers were immersed in the substrate at a distance
of 8\,$\mu$m from the top and bottom surfaces and with a spacing
\textit{s}=24\,$\mu$m between two layers. The single mesh layers
are periodic structures of square grids patterns. In this design $
\textit{a/g}=0.14$ with \textit{g}=25.4\,$\mu$m and
\textit{a}=3.556\,$\mu$m. Once the 40\,$\mu$m multi-layer
structure is assembled it is then fused by hot-pressing the layers
at high temperature near the polypropylene melting point
(160$^{\circ}$\,C). The temperature is such that the polypropylene
layers are joined but is insufficient to allow the polymer chains
to flow in a liquid-like manner. Expansion due to relaxation when
hot pressing the layers is avoided by maintaining high pressure in
the press and through the absence of soft materials in the stack.
Thicknesses measurements made after hot pressing show that the
layer thicknesses are maintained to better than $\pm 1$\,$\mu$m.

\begin{figure}[h!]\label{Fig-6}
\centering
\includegraphics[width=12cm,height=6.5cm,angle=0]{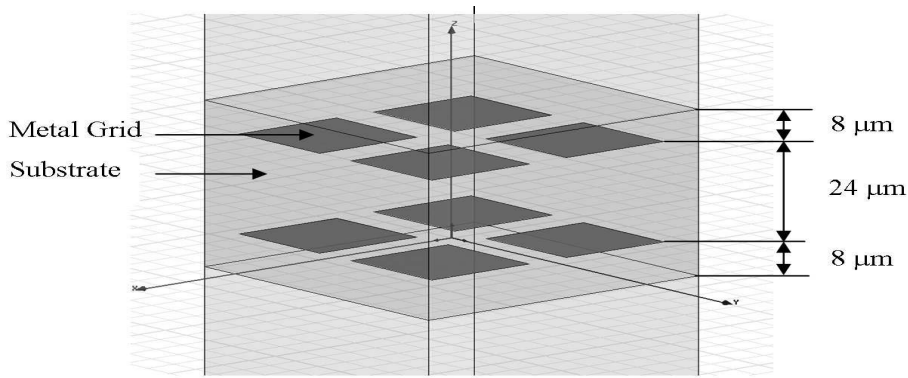}
\caption{Prototype ARC metal mesh model.}
\end{figure}

The artificial material created by this process has the
consistency of a solid plastic film which can be easily handled
and cut to size to suit requirements. Bonding of this material to
the quartz crystal is again performed through a cycle of thermal
heating with applied pressure using a thin layer of polymer based
glue (thickness $<<$ wavelength). Finally, the outer layers of
porous PTFE are again glued at the extremes of the stack (top
layer in Figure 7)with identical procedure (the two procedures can
in fact be performed in one single press cycle).

\begin{figure}[h!]\label{Fig-7}
\centering
\includegraphics[width=12cm,height=9cm,angle=0]{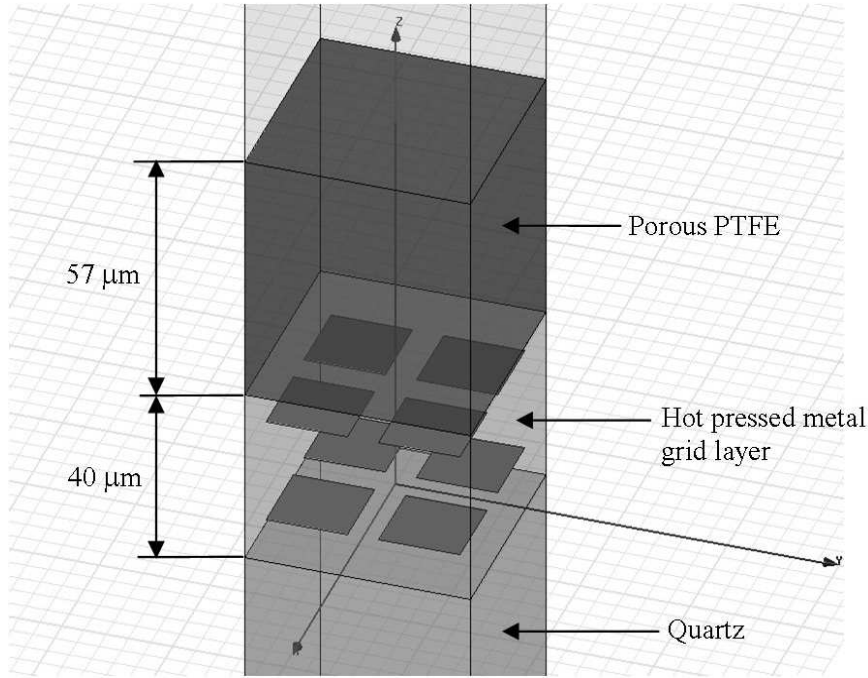}
\caption{Full model ARC on both sides of quartz substrate.}
\end{figure}

The ARC layers are symmetrical with respect to the central crystal
in order to taper the impedance that a given wavelength of
radiation experiences when entering and subsequently exiting the
crystal.

\subsection{HFSS prediction for anti-reflection coating}

Simulations for both the prototype ARC on its own and the
completed quartz plate with AR coatings have been made. The
transmission for the prototype ARC layer is shown in Figure 8. The
refractive index was calculated based on transmission as described
in section 2.3 to be 1.65 in average. Figure 9 shows the
comparison between the uncoated quartz substrate, the AR-coated
substrate and the simulated transmission expected. The efficiency
of the coating yielding high transmission over the frequency band
from 18.6--52\,cm$^{-1}$ (560--1560\,GHz) is obvious.

\begin{figure}[h!]\label{Fig-8}
\centering
\includegraphics[width=12cm,height=8cm,angle=0]{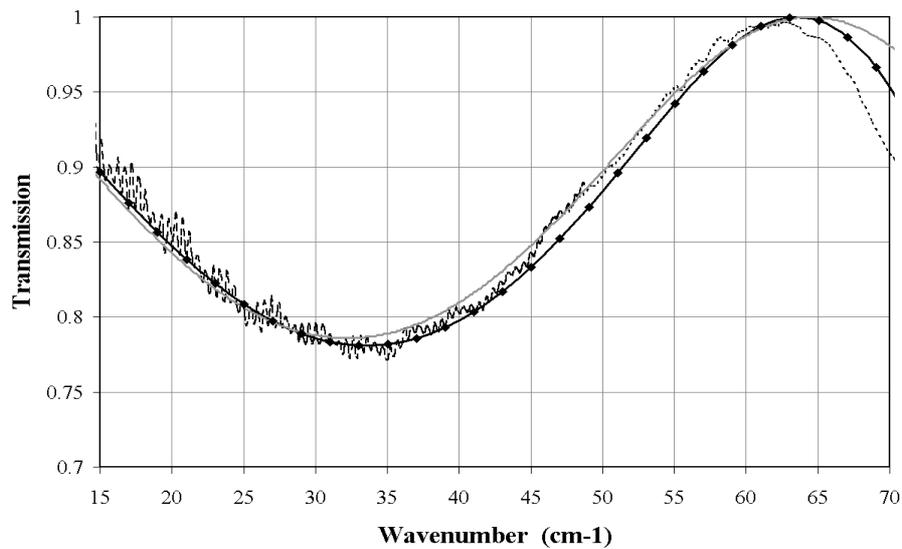}
\caption{Transmission for ARC metal mesh layer: measurements
(black dotted line), HFSS simulation (black line with diamond
symbols) and Fabry-Perot model (grey line).}
\end{figure}

\begin{figure}[h!]\label{Fig-9}
\centering
\includegraphics[width=12cm,height=8cm,angle=0]{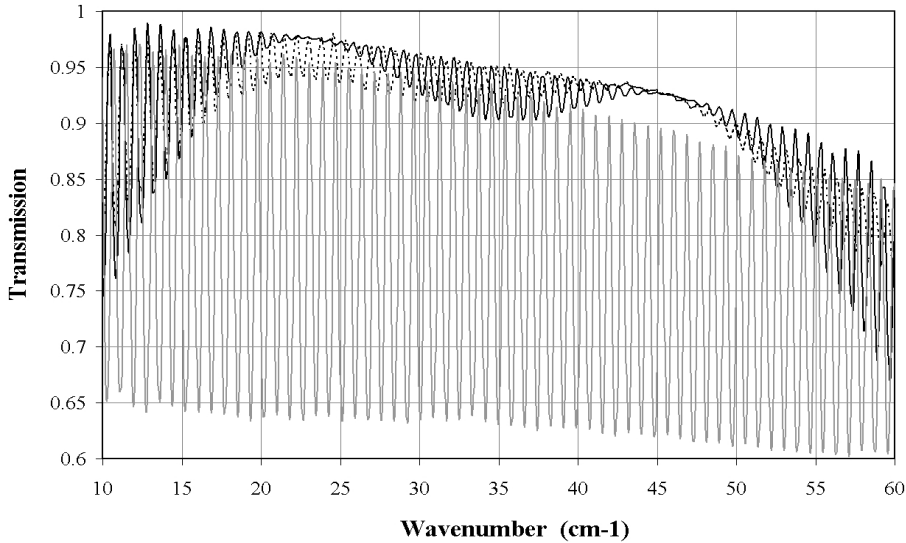}
\caption{Transmission of the complete ARC quartz plate.  Lower
grey line is measured data of the uncoated quartz substrate, the
black line is the HFSS simulation of the complete ARC quartz plate
and the black dotted line is the measured data of the complete ARC
quartz plate.}
\end{figure}

\section{Measurements and Data Analysis}

As a stand-alone material, the embedded grids in polypropylene
will have a complex impedance with the real part mimicking a
behaviour analogous to that of a plane parallel dielectric slab.
We therefore subjected the material to the same spectral tests
that we would have imposed on a thin dielectric slab to determine
its refractive and absorptive properties.

The material has also been measured when used as a tuned
anti-reflection coating element.  As reported above for this
measurement it was hot pressed on both sides of a quartz plate. We
were careful to choose grids with a period of 25.4\,$\mu$m to
ensure that diffraction effects would be negligible at the
frequencies for which the ARC was optimised. However, we expected
to observe spectral features resulting from the complex impedance
nature of the resonant structures which would distinguish the
measured spectral behaviour from that of a normal dielectric with
constant index of refraction.

\subsection{Spectral measurements of the material.}

To verify the optical properties of the artificial dielectric
alone and with the crystal, we placed the material in the
\textit{f}/3 beam of a rapid scanning polarizing Fourier Transform
Spectrometer (FTS). The spectral features associated with the
source (a Hg-arc lamp) together with the spectral efficiency of
the detector (a 4 Kelvin Ge bolometer) were removed by performing
the ratio with a background spectrum recorded with the sample
removed. Three sets of 10 scans each were performed in order to
obtain acceptable signal to noise for a spectral resolution of
0.25\,cm$^{-1}$ and limit phase errors in the Fourier transform
process. Interferograms in each set were averaged and then Fourier
Transformed. The resulting three spectra have then been averaged
before ratioing against the background spectrum. The final
transmission spectrum is shown in Figure 8. By repeating this
measurement at various orientation of the mesh axises with respect
to the polarization direction we confirmed there is no
polarization sensitivity observable above 1\,\% level as expected.

From the peak of transmission at 62\,cm$^{-1}$ we can infer that
absorption loss due to the polypropylene, which is usually a
monotonic function of frequency, is negligible.  Data for the
absorption coefficient of Tucker and Ade \cite{23} indicate that
the expected loss would be only 0.2\,\%. With this consideration
we can expect the transmission of an equivalent dielectric slab to
be given by equation (1). By inverting this equation whilst fixing
the physical thickness $t$ of the material we can numerically
solve for each value of $\tilde{\sigma}$ to obtain a value
$n(\tilde{\sigma})$. Figure 10 shows the residual mismatch of the
FTS data with the ideal Fabry-Perot behaviour. The contour levels
are the percentage error on the transmission mismatch
(measured/ideal). The thick line in the clear contour region shows
the best fit over the spectral interval 20--40\,cm$^{-1}$ for an
assumed linear function of frequency. The linear fit parameters
are:
\begin{eqnarray}
    \textbf{n} & = & n_0 + n_1\,\tilde{\nu} \\
    n_0 & = & 1.637 \\
    n_1 & = & 2.07 \times 10^{-3}
\end{eqnarray}

\begin{figure}[h!]\label{Fig-10}
\centering
\includegraphics[width=12cm,angle=0]{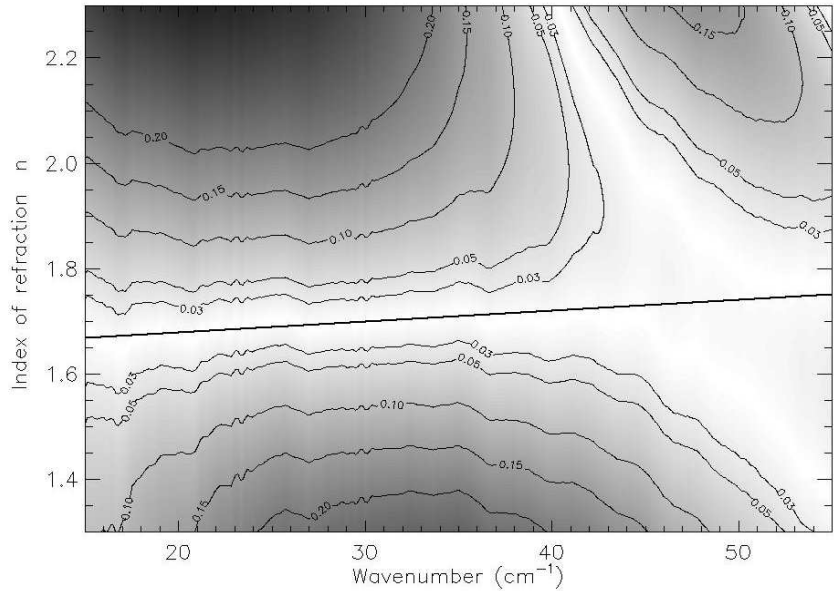}
\caption{Residual contour of measured transmission vs expected
ideal behaviour of an ideal dielectric slab. The thick line in the
clear region denotes the best fit linear function of frequency.}
\end{figure}

The small ripple artifacts which are visible in the leftmost
region of the contour, arise from standing waves in the FTS sample
box between the sample and the output polariser, and hence have
been ignored in the fitting procedure.

Using these values as initial conditions for a simulation of the
complete stack we can then proceed to verify that its employment
as an anti-reflection coating element is consistent with its
stand-alone properties.

\subsection{Spectral measurements of coated quartz substrate.}

As previously stated, the end goal is to produce a broadband
anti-reflection coating to remove reflection loss from high
refractive index crystalline components. The complete ARC
structure as described above was measured in the identical setup
as the stand-alone material.  Figure 9 presents these data which
can now be compared with the spectral performance as simulated at
a number of frequencies using HFSS.

First, we note that in general there is good agreement between the
envelopes of the fringes. This denotes that the material is acting
as a good anti-reflection coating over the desired band. However,
some minor discrepancies are also present which question either
the set of input parameters for the simulation or our
understanding of the effects.

Having manufactured a large number of anti-reflection coated
windows and filters with different techniques, we are aware that
some of the parameters can slightly change due to the heat-bonding
process in our press. Indeed some bonding materials change when
heated. Porous PTFE can be pressed to a smaller thickness and
being porous can absorb some of the polyethylene glue, raising the
equivalent index of refraction by a small amount and changing its
final thickness. Polypropylene on the other hand tends to relax
and expand if there is not sufficient pressure placed on it.
Crystals clearly will not be deformed at these temperatures and
pressures.

Performing HFSS simulations with different values of parameters at
multiple frequencies proves to be very time consuming, so we have
adopted a transmission line based model \cite{22} to perform a
Monte-Carlo simulation cycling through some of the input
parameters (thickness of the coating layers and indexes of
refraction to find the solution which closely matches our data set
and check for potential manufacturing effects.

In order to justify the results we first tested the consistency of
the two simulation/modelling packages by running the transmission
line model on the nominal input-parameters. In Figure 11 we can
clearly see an excellent agreement from low to high-frequencies
with the two best-matching regions in 20--25 and
42--46\,cm$^{-1}$. The loss of relative precision in the HFSS
simulation at the higher frequencies is potentially the cause of
the mismatch occurring there.

\begin{figure}[h!]\label{Fig-11}
\centering
\includegraphics[width=12cm,height=6.5cm,angle=0]{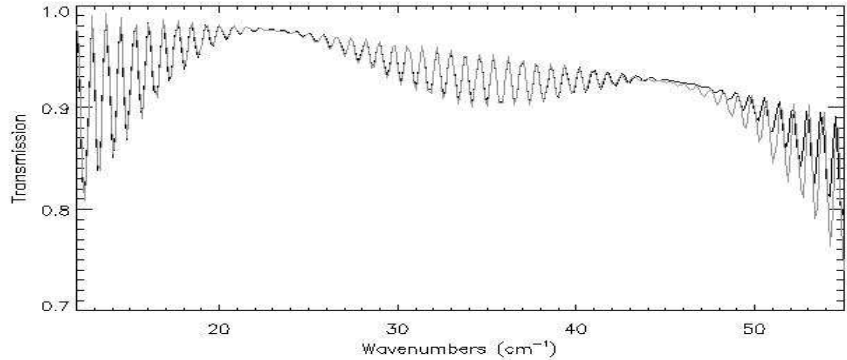}
\caption{Expected match of the scattering matrix model (grey) with
the HFSS simulation (black dotted).}
\end{figure}

\begin{figure}[h!]\label{Fig-12}
\centering
\includegraphics[width=12cm,height=6.5cm,angle=0]{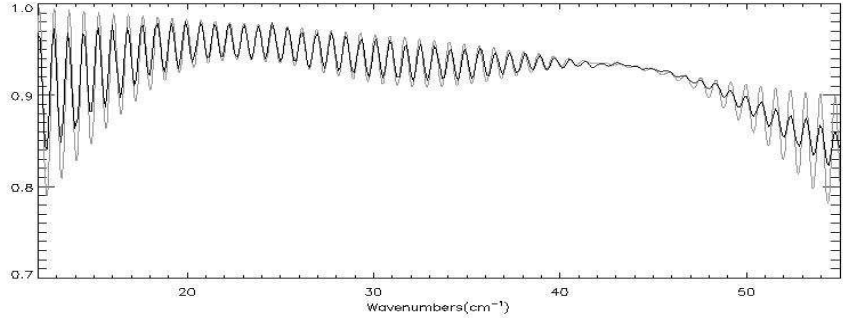}
\caption{Transmission spectrum of the stack
(PPTFE-ADM-Quartz-ADM-PPTFE). The black line is the best fit of a
scattering matrix model with varying optical constants to fit the
behavior of the ADM as a function of frequency.}
\end{figure}

The Monte-Carlo parameter fit results and minimum square residuals
are given in Table 1 for the best fit of the data shown in Figure
12. This fit, performed over the region 15--55\,cm$^{-1}$, yields
a good match with the following caveats regarding the numbers in
Table 1:

\begin{itemize}
 \item \underline{Mupor \cite{25}:} Porous PTFE. The increase in the mupor index and its decrease in average thickness is consistent with compression of a porous material during bonding.
 \item \underline{ADM - t:} The thickness of the Artificial Dielectric Material (ADM). The values of thickness slightly larger than expected could point to thermal relaxation of the polypropylene layers. This phenomenon has been observed in direct coating of quartz windows. This tendency of thin layers to shrink in diameter and grow in thickness of a few percent can occur when the pressure applied is not sufficient or other less dense materials in the stack give way.
 \item \underline{ADM - n:} The refractive index of the Artificial Dielectric Material (ADM). The values of index of refraction are slightly different from the recovered functional values found previously ($n_0=1.62$ and $n_1=2\,\times\,10^{-3}$) but it is important to note that in the interval of frequencies considered, the two functional dependencies are not too dissimilar.
 \item \underline{Top-bottom asymmetry :} Generally we would not expect a marked asymmetry between top and bottom coating layer thicknesses. However, we note that in the vacuum oven the top plate of the press is only conductively heated through the stainless steel press screws. A temperature difference of a few degrees will affect the tendency of materials to change their thickness subject to pressure.
 \item \underline{Quartz n :} The values of quartz index obtained are slightly different (smaller) than those in literature \cite{26} by a scaling factor $\sim 1/1.007$. This is expected as the standing wave pattern in the fitted data does not vary with a constant product of $n$ and the quartz thickness.
 To avoid additional standing waves in the measuring setup, the sample was tilted by $\hat{i}=\simeq 10^{\circ}\pm 5^{\circ}$ and this would affect the thickness travelled by
 $1/{\cos[\sin^{-1}(n\cdot \sin \hat{i})]} \simeq 1/(1+\epsilon)$ with $\epsilon \subset [1,8]\cdot 10^{-3}$. By forcing the nominal thickness the fitted value of the index of refraction compensates accordingly.

\end{itemize}

\begin{table}[h!]
{\bf \caption{Simulation results.}} List of parameters (fixed and
variable) in the simulation of the coated stack. It is important
to specify that due to the nature of the fit and the different
parameter intervals, the errors reported are only indicative of
the dependence of the result from that parameter (i.e. how much
does that variable influence the result) and not a simple standard
deviation of a gaussian distribution. The two errors are
respectively indicating an increase of (1)=10\,\% and (2)=\,100\%
in the residual of the least square quadratic fit when applied to
the specific parameter. $(*)$ The quartz value was measured on the
substrate with a precision micrometer and the error resulting in
the repeatability of the measurement across the sample.

\begin{center}
\begin{tabular}{lccccc}\hline%{|l{3.5in}|c{1.5in}|c{1.0in}|c{1.0in}|}\hline
    {\bf Parameter} & $\#$ & Value & Error$^{(1)} (\pm)$ & Error$^{(2)} (\pm)$ & Units \\
    \hline
    Mu-por Index of Refraction     & $n_{\mu por}$   & $1.375$   & $0.01$  & $0.03$ & \\
    \hline
    Mu-por Thickness (Top)         & $t^t_{\mu por}$ & $41.9$    & $1.5$   & $4$    & ($\mu$m) \\
    \hline
    Mu-por Thickness (Bottom)      & $t^b_{\mu por}$ & $44.1$    & $1.7$   & $4$    & ($\mu$m) \\
    \hline
    ADM Index of Refraction        & $n_0$           & $1.592$   & $0.014$ & $0.04$ &  \\
    \hline
    ADM Index of Refraction        & $n_1$           & $7.5e-3$  & $5e-4$  & $1.4e-3$ & \\
    \hline
    ADM Thickness (Top)            & $t^t_{ADM}$     & $42$      & $^{+2}_{-1}$ & $^{+5}_{-2}$ & ($\mu$m) \\
    \hline
    ADM Thickness (Bottom)         & $t^b_{ADM}$     & 45        & $^{+2}_{-1}$ & $5$ & ($\mu$m) \\
    \hline
    Quartz Index of Refraction     & $n_0$           & $2.095$   & $0.002$ & $0.005 $ & \\
    \hline
    Quartz Index of Refraction     & $n_1$           & $4e-6$    & $4e-5$ & $1.4e-4$  & \\
    \hline
    Quartz Index of Refraction     & $n_2$           & $2e-7$    & $1e-6$ & $4e-6$ & \\
    \hline
    Quartz Loss Tangent            & $\alpha_1$      & $4.1e-5$    & $7e-6$ & $2e-5$ & \\
    \hline
    Quartz Loss Tangent            & $\alpha_2$      & $4.0e-6$  & $2e-7$ & $6e-7$ & \\
    \hline
    Quartz Thickness (*)           & $d$(fixed)      & $2.890 $  & $-$  & $0.003$ & $(mm)$\\
    \hline
\end{tabular}
\end{center}
\end{table}

\section{Conclusions}

We have shown through modelling and experimental verification that
artificial dielectrics with refractive indexes above that of the
base material can be obtained over a broad spectral band by fusing
in layers of metal mesh.  The resultant refractive index can be
easily controlled by adjusting the geometrical parameters of the
meshes and the spacing between them in the new material.

We have also demonstrated the viability of this metamaterial as an
element in producing a broad band anti-reflection coating for a
quartz crystal.  Again modelling and measurement results are in
good agreement showing that we have good control of this process.
As a further test of the robustness of the coated layers we
dropped the sample into liquid nitrogen (77K) and warmed to room
temperatures five times with no discernable signs of the coating
peeling of fracturing.

We believe that this new approach will assist in the manufacture
of many other optical components in the FIR to millimetre
wavelength range.  These include flat lenses, birefringent
components for waveplates (using 1-D grid structures) and
dispersive components.

\bibliographystyle{unsrt}

\begin{thebibliography}{}

\end{thebibliography}


\begin{thebibliography}{1000}\label{thebibliography}

\bibitem{1}Ade P.A.R., Pisano G., Tucker C., Weaver S., ``A Review of Metal Mesh Filters'', Proceedings of the SPIE, Vol. 6275, 62750U (2006).

\bibitem{2}Ulrich R., ``Interference Filters for The Far Infrared'', Applied Optics, Vol. 7 , No. 10, October 1968.

\bibitem{3}Kuznetsov S.A., Kubarev V.V., Kalinin P.V., Goldenberg B.G., Eliseev V.V., Petrova E.V., and Vinokurov N.A., ``Development of Metal Mesh Based Quasi-Optical Selective Components and Their Applications in High-Power Experiments at Novosibirsk Terahertz Fel'', Proceedings of FEL 2007, pp. 89--92, Novosibirsk, Russia, 2007.

\bibitem{4}Naylor D.A., Gom B.G., Schofield I. S., Tompkins G. J., Davis G.R., ``Mach-Zehnder Fourier Transform Spectrometer for Astronomical Spectroscopy at Submillimeter Wavelengths'', SPIE, Millimeter and Submillimeter Detectors for Astronomy 4855, 540-551, 2003.

\bibitem{5}Lamarre, J. M.; Puget, J. L.; Bouchet, F.; Ade, P. A. R.; Benoit, A.; Bernard, J. P.; Bock, J.; de Bernardis, P.; Charra, J.; Couchot, F.; Delabrouille, J.; Efstathiou, G.; Giard, M.; Guyot, G.; Lange, A.; Maffei, B.; Murphy, A.; Pajot, F.; Piat, M.; Ristorcelli, I.; Santos, D.; Sudiwala, R.; Sygnet, J. F.; Torre, J. P.; Yurchenko, V.; Yvon, D. ``The Planck High Frequency Instrument, a third generation CMB experiment, and a full sky submillimeter survey'', New Astronomy Reviews, Volume 47, Issue 11-12, p. 1017-1024, 2004.

\bibitem{6}Fowler J.W., Niemack M. D., Dicker S. R., Aboobaker A. M., Ade P. A. R., Battistelli E. S., Devlin M. J., Fisher R.P., Halpern M., Hargrave P.C., Hincks A.D.,Kaul M., Klein J., Lau J. M., Limon M., Marriage T. A., Mauskopf P.D., Page L., Staggs S.T., Swetz D. S., Switzer E. R., Thornton R. J., Tucker C. E., ``Optical Design of the Atacama Cosmology Telescope and the Millimeter Bolometric Array Camera'', Applied Optics, 46, Issue 17, 3444-3454, 2007

\bibitem{7}Pascale, E.; Ade, P. A. R.; Bock, J. J.; Chapin, E. L.; Chung, J.; Devlin, M. J.; Dicker, S.; Griffin, M.; Gundersen, J. O.; Halpern, M.; Hargrave, P. C.; Hughes, D. H.; Klein, J.; MacTavish, C. J.; Marsden, G.; Martin, P. G.; Martin, T. G.; Mauskopf, P.; Netterfield, C. B.; Olmi, L.; Patanchon, G.; Rex, M.; Scott, D.; Semisch, C.; Thomas, N.; Truch, M. D. P.; Tucker, C.; Tucker, G. S.; Viero, M. P.; Wiebe, D. V., ``The Balloon-borne Large Aperture Submillimeter Telescope: BLAST'', Astrophys. J., 681, Issue 1, 400-414, 2008

\bibitem{8}Griffin, M.; Ade, P.; André, Ph.; Baluteau, J.-P.; Bock, J.; Franceschini, A.; Gear, W.; Glenn, J.; Huang, M.; King, K.; Lellouch, E.; Naylor, D.; Oliver, S.; Olofsson, G.; Page, M.; Perez-Fournon, I.; Rowan-Robinson, M.; Saraceno, P.; Sawyer, E.; Swinyard, B.; Vigroux, L.; Wright, G.; Zavagno, A., ``The SPIRE Instrument'', EAS Publications Series, 34, 33-42, 2009.

\bibitem{9}Ruhl, John; Ade, Peter A. R.; Carlstrom, John E.; Cho, Hsiao-Mei; Crawford, Thomas; Dobbs, Matt; Greer, Chris H.; Halverson, Nils w.; Holzapfel, William L.; Lanting, Trevor M.; Lee, Adrian T.; Leitch, Erik M.; Leong, Jon; Lu, Wenyang; Lueker, Martin; Mehl, Jared; Meyer, Stephan S.; Mohr, Joe J.; Padin, Steve; Plagge, T.; Pryke, Clem; Runyan, Marcus C.; Schwan, Dan; Sharp, M. K.; Spieler, Helmuth; Staniszewski, Zak; Stark, Antony A., ``The South Pole Telescope'', SPIE, Volume 5498, pp. 11-29,

\bibitem{10}Pisano G., Savini G., Ade P.A.R., Haynes V., ``Metal-mesh Achromatic Half-wave Plate for Use at Submillimeter Wavelengths'', Applied Optics, Vol. 47, No. 33. 20 November 2008.

\bibitem{11}http://www.ansoft.co.uk/products/hf/hfss/.

\bibitem{12}Marcuvitz N., ``Waveguide Handbook'', M.I.T. Rad. Lab. Ser., McGraw-Hill, 280-290, 1951.

\bibitem{13}Compton R., Rutledge D. B., ``Approximation Techniques for Planar Periodic Structures'', IEEE Transactions on Microwave Theory And Techniques, Vol. MTT-33, No.10, October 1985.

\bibitem{14}Chen C.C., ``Transmission of Microwave Through Perforated Flat Plates of Finite Thickness'', IEEE Transactions on Microwave Theory And Techniques, Vol. MTT-21, No.1, January 1973.

\bibitem{15}Rubin B., ``Scattering from a Periodic Array of Apertures or Plates Where the Conductors Have Arbitrary Shape, Thickness, and Resistivity'', IEEE Transactions On Antennas And Propagation, Vol. AP-34, NO.11, November 1986.

\bibitem{16}Shen J.T., Catrysse P.B., and Fan Shanhui., ``Mechanism for Designing Metallic Metamaterials with a High Index Refraction'', Physical Review Letters. PRL 94, 197401, 2005.

\bibitem{17}Smith D.R., Schultz S., Markos P., and Soukoulis Markos P., ``Determination of Effective Permittivity and Permeability of Metamaterials from Reflection and Transmission Coefficients'', Physical Review B, Vol. 65, 195104, 2002.

\bibitem{18}Hecht E., ``Optics'', ISBN 0-321-18878-0, Addison Wesley, 1301 Sansone St., San Francisco, CA94111, 2002.

\bibitem{19}Pisano G., Savini G., Ade P.A.R., Haynes V., and Gear W. K., ``Achromatic Half-Wave Plate for Submillimeter Instrumentations in Cosmic Microwave Background Astronomy: Experimental Characterization',' Applied Optics, Vol. 45, No. 27, September 2006.

\bibitem{20}Savini G., Ade P.A.R., House J., Pisano G., Haynes V., Bastien P., ``Recovering The Frequency Dependent Modulation Function of The Achromatic Half-wave Plate for POL-2: The SCUBA-2 Polarimeter'', Applied Optics, Vol. 48, pp. 2006--2013, 2009.

\bibitem{21}Rogers Corporation supply TMM high frequency laminate materials:
www.rogerscorp.com

\bibitem{22}Savini G., Pisano G., and Ade P.A.R., ``Achromatic Half-Wave Plate for Submillimeter Instruments in Cosmic Microwave Background Astronomy: Modeling and Simulation'', Applied Optics, Vol. 45, No. 35, August 2006.

\bibitem{23} Tucker C. and Ade P.A.R., ``Thermal filtering for large aperture cryogenic detector arrays'', Proceedings of the SPIE, Vol. 6275, pp. 62750T, 2006.

\bibitem{24}Tucker C. and Ade P.A.R., ``Metal-Mesh Filters for THz Applications'', Infrared and Millimeter Waves, 2007 and the 2007 15th International Conference on Terahertz Electronics, IRMMW-THz Joint 32nd International Conference, pp. 973-975, 2007.

\bibitem{25}http://www.porex.com/porous.cfm

\bibitem{26} Russell E.E. and  Bell E.E., ``Measurement of the optical constants of crystal quartz in the far infrared with the asymmetric Fourier-transform method'', Journal of the Optical Society of America, vol. 57, issue 3, p.341

\end{thebibliography}

\end{document}